# Comment on

# Quantum Theory Looks at Time Travel:  D.M.Greenberger &K.Svozil

John S. Wykes

*Ashleigh Coach House, Ashleigh Drive*
*Chellaston, Derby DE73 1RG, UK*

Abstract

It is claimed in the above paper that, if time travel were possible, quantum propagation would prevent classic time-travel paradoxes such as killing one's grandfather, by establishing consistent time loops; an example circuit is used to demonstrate such a loop. It is argued here that established loops are not the framework in which the classic paradoxes arise; rather they arise via the establishment of a concrete initial history in which no disturbing time travel is allowed and then disturbing that history via the launch of the time traveller. It is shown that, operated in this two-pass fashion, if the first forward evolution of the circuit produces a definite triggering of a backwards time travelling state, the re-evolution thereby engendered may be organised so as to prohibit the triggering of this state, thereby creating a classical time travel paradox.

---

Electronic address: john.wykes@talk21.com

In 'Quantum Theory Looks at Time Travel' [1], (QTLTT), Greenberger and Svozil set up a quantum time travel loop and claim to show that nothing corresponding to the classical time travel paradoxes, such as shooting one's father prior to one's conception, can occur.

However the analysis of a fully established time travel loop, evolving unitarily without state collapsing measurement processes, is not one in which such paradoxes are normally set. Rather the paradox is expressed in terms on an initial evolution through a time interval without the complicating presence of a time travelling channel (in the Father paradox this is the pass in which one is conceived, born and grows to adulthood free from interference from one's time travelled self), which is then followed by the use of a time travel channel to disturb the system at some past time in such a way as to eliminate the possibility of one's own conception and so create the paradox when the system re-evolves under the disturbed condition yielding a contradictory result.

The QTLTT circuit consists of two identical beam splitters, the first of which is encountered from the right by an input state $\psi$ at time $t_1$; the two resulting states, $\psi_1$ and $\psi_2$, propagate in left and right hand channels respectively, with associated propagators $G_1$ and $G_2$, until they are re-combined by encountering opposite sides of the second beam splitter at time $t_2$; the right and left outputs of this splitter are labelled $\psi_3$ and $\psi_4$ respectively; the left output then encounters a time-travelling channel that propagates it back to form the left input of the first beam splitter at the previous time $t_1$: - see Figure 3 of QTLTT, where the time travel channel is drawn as an additional leftmost feedback arm of the circuit, and has propagator M.

To correspond to the conditions used when discussing classical time travel paradoxes one should initially omit the leftmost arm in which time travel is happening (or equivalently have the propagator in that channel, M, set to zero) and allow the system to evolve as normal to create the first of what will be the two contradictory conclusions necessary for a paradox.

Using the QTLTT notation, we find that, without this backwards propagation arm, the circuit yields for the two outputs from the second beam-splitters at time $t = t_2$

$$\psi_3(t_2) = (\alpha^2 G_1 - \beta^2 G_2)\psi(t_1) \qquad (a)$$

$$\psi_4(t_2) = -i\alpha\beta(G_1+G_2)\psi(t_1) \qquad (b)$$

where $\alpha$ and $\beta$ are the transmission and reflection amplitudes, respectively, of the beam splitters. Adopting the values given by QTLTT of $\alpha,\beta = 1/\sqrt{2}$, $G_1 = -iG_2 = G$, we see that this circuit simply produces propagation from $t_1$ to $t_2$ via G and then a 50:50 split of the input state into the two output channels i.e.

$$\psi_3(t_2) = \psi_4(t_2) = [(1-i)/2]G\psi(t_1). \qquad (c)$$

Subsequent measurement would show one or other output channel being occupied with a 0.5 probability of each.

As stated above a classical paradox requires upon two occurrences contradicting each other. For this first evolution of the system to count as 'an occurrence' the state must undergo the process of reduction ('collapse') which, via environmental decohering processes for example, stamps its presence on the future in a locally irreversible manner. We now allow that, in the event of the collapse of the output into the left-hand output channel, the availability of the time travel channel back to $t_1$ is triggered. The time travel channel now launches a state propagating backwards in time. Let this launched state be $\psi_T(t_2)$. This is transported, via the propagator M, back from $t_2$ to $t_1$ down the leftmost arm of the QTLTT circuit, to be input to the left hand side of the first beam splitter. It is then the second evolution through the forward channels of the system, with inputs $\psi(t_1)$ on the right (as before) and $\psi_T(t_1) = M.\psi_T(t_2)$ on the left, that can create the paradox.

Following these inputs through the circuit from $t_1$ to $t_2$ we find that the two output states now each have an additional term created by the input of the time travelled state at $t_1$,

$$\psi_3(t_2) = (\alpha^2 G_1 - \beta^2 G_2)\psi(t_1) - i\alpha\beta(G_1+G_2)M\psi_T(t_2) \tag{d}$$

$$\psi_4(t_2) = - i\alpha\beta(G_1+G_2)\psi(t_1) + (\alpha^2 G_2 - \beta^2 G_1)M\psi_T(t_2) \tag{e}$$

If we can now choose M such that

$$M\psi_T(t_2) = \psi(t_1), \tag{f}$$

which one may see as a coherence condition on the triggered time-travelling channel, then, using the same values of $\alpha,\beta$, $G_1$ and $G_2$ we used to arrive at (c), we find

$$\psi_3(t_2) = (1- i)G\psi(t_1), \tag{g}$$

$$\&\quad \psi_4(t_2) = 0. \tag{h}$$

This shows that the time travelling state launched at $t_2$ has acted as a classic father-killer by cancelling the output state that launched it and has thereby created a paradox at $t_2$: the definite left output of the first pass has ensured, via the backward propagation of $\psi_T$, that the output of the second pass must be into the right channel only.

If quantum mechanics is to prevent such paradoxes under such circumstances then the coherence condition (f) on the time-travelling state must be disallowed. Although the decohering process associated with creating the first occurrence in the left output may, in general, be such as to produce a random phase addition to $M\psi_T(t_1)$ with respect to $\psi(t_1)$ and thereby produce something less than a total contradiction, there seems to be no reason in principle that the full contradiction described above could not occur.

References
[1]     Quantum Theory Looks at Time Travel: Daniel M. Greenberger and Karl Svozil,  arXiv:quant-ph/0506027v2